\documentclass[twoside,slac_one]{revtex4}
\usepackage{graphicx}
\usepackage{fancyhdr}
\usepackage{amsmath} 
\usepackage{bm}
\usepackage{amsxtra}
\usepackage{amssymb}
\usepackage{amsthm}
\usepackage{latexsym}
\usepackage{lscape}

\begin{document}

\title{Search for Pair Production of Scalar Leptoquarks}

%

\author{E. Barberis for the CMS Collaboration}
\affiliation{Department of Physics, Northeastern University, Boston, MA, USA}

\begin{abstract}
This article describes the search for pair production of scalar leptoquarks performed by the CMS collaboration
using the data from the 2010 proton-proton run of the Large Hadron Collider at a center of mass energy of 7 TeV. The following
 final state signatures from the decay of a pair of scalar leptoquarks are described in here: the di-lepton plus jets final state
 (where the leptons are either both electrons or both muons) and the lepton plus jets and transverse missing energy final state
 (where the lepton is an electron).
\end{abstract}

\maketitle

\thispagestyle{fancy}


\section{Introduction}
Leptoquarks (LQ) are hypothetical particles that are predicted by many extensions~\cite{mBRW,techni1,techni2,techni3,superstring} of the Standard Model of particle physics (SM), such as Grand
Unification Theory, Technicolor and Composite models. They carry both baryon and lepton numbers and thus couple to a
lepton and a quark. They carry fractional electric charge, they are color triples, and can have either zero or one unit of spin (i.e. can be either scalar or vector particles). Existing experimental limits on lepton number violation, Flavor Changing Neutral Current, proton decay, and
other rare processes favor three generations of leptoquarks, with no inter-generational mixing.

The production and decay of LQs are characterized by: the mass of the LQ particle (M$_{LQ}$), its decay branching ratio into a lepton and a quark (usually denoted as $\beta$) and the Yukawa coupling at the LQ-lepton-quark vertex ($\lambda$). At hadron colliders, leptoquarks are mainly produced in pairs, via gluon-gluon fusion and quark anti-quark annihilation. The dominant pair production mechanisms do not depend strongly on $\lambda$ and the single production of leptoquarks does not become significant (and it does not in any case invalidate the search results presented here) in the range of LQ masses probed with the 2010 LHC data.

The final state event signatures from the decay of pair produced LQs can be classified as: dilepton and jets (both LQ and anti-LQ decay into a charged lepton and a quark); lepton, missing transverse energy and a jet (one LQ decays into a charged lepton and a quark, while the other decays into a neutrino and a quark); and missing transverse energy and jets (both LQ and anti-LQ decay into neutrinos and quarks). The three signatures have branching ratios of $\beta^2$, 2$\beta(1-\beta)$, and $(1-\beta)^2$, respectively. The charged leptons can be either electrons, muons, or tau leptons, corresponding to the three generation of LQs. Only electrons and muons are considered here.

\section{Search for Pair-Production of Scalar Leptoquarks with the CMS Detector}
Searches for first and second generation pair-produced scalar LQs were performed and published on 34-36 pb$^{-1}$ of 7 TeV proton-proton collision
data recorded by the CMS detector~\cite{CMSJinst} during the 2010 LHC run. First generation results include searches both in the dilepton and jets final state ($ee jj$), and in the lepton, missing transverse energy and jets final state ($e\nu jj$). Second generation result consist of a search in the dilepton and jets final state ($\mu\mu jj$). The $ee jj$ and $e\nu jj$ results were combined to attain the best possible exclusion reach in all of the parameter space of the first generation pair produced scalar LQ search, i.e. ($\beta$ and $M_{LQ}$).

The analysis in all of the channels aims at identifying the existence of new heavy particles by establishing an excess of events characteristic of the decay of heavy objects. The analysis starts with using either a single or a double lepton trigger path, which is robust and very efficient. An event signature preselection isolates events with high transverse momentum final state objects (two or more isolated leptons and two or more jets, or one isolated lepton, two or more jets, and significant missing transverse energy indicative of the presence of a neutrino). Event variables are identified to effectively separate a possible LQ signal from standard model backgrounds (these are the $S_T$, $M(ll)$, and $M_T(l\nu)$ variables
described below) and lower thresholds are placed on these variables at preselection level. Backgrounds from standard model sources are estimated and first compared with data at the preselection level. Major sources of background are $Z/\gamma^*$+jets and $W+$jets processes and $t\bar{t}$ (with single top production, diboson processes, and QCD multijet processes being smaller contributions). Major backgrounds are either directly determined from data control samples or determined with Monte Carlo samples (therefore using kinematic shape information from the MC) but normalized to data in selected control regions.

A cut-based variable approach is used for the final selection, where the selection is optimized for different
LQ mass hypotheses by minimizing the expected upper limit on the LQ cross section in the absence of an observed signal
using a Bayesian approach~\cite{bayes1, bayes}. The three variables used in the optimization are: the scalar sum of the final objects transverse momenta, $S_T$;
the invariant mass of the dilepton pair, $M_{ll}$, for the $ee jj$ and $\mu\mu jj$ channels; and the transverse mass, $M_T(l\nu)$, of the lepton and
neutrino in the $e\nu jj$ analysis.

After final selection, the data are well described by the standard model background predictions. In the absence of an observed signal, an upper limit on the LQ cross section is set using a Bayesian method~\cite{bayes1, bayes}
with a flat signal prior. A lognormal prior is used to integrate over the nuisance parameters. Using Poisson statistics,
a 95\% confidence level (C.L.) upper limit is obtained on the LQ pair-production cross-section times branching ratio as a function of LQ mass. Comparing with the NLO predictions~\cite{NLO} for the scalar LQ pair production cross section a lower limit on $M_{LQ}$ is determined for $\beta = 1$ and $\beta = 0.5$ (electron only, for the latter). In the electron channel, the $ee jj$ and $e\nu jj$ channels are combined to further maximize the exclusion in $\beta$ and $M_{LQ}$, specially for the case of $\beta = 0.5$, where the $ee jj$ adds to the sensitivity of the $e\nu jj$ channel.

\section{$ee jj$ and $\mu\mu jj$ channels}

\begin{figure}[ht]
\centering
\includegraphics[width=80mm]{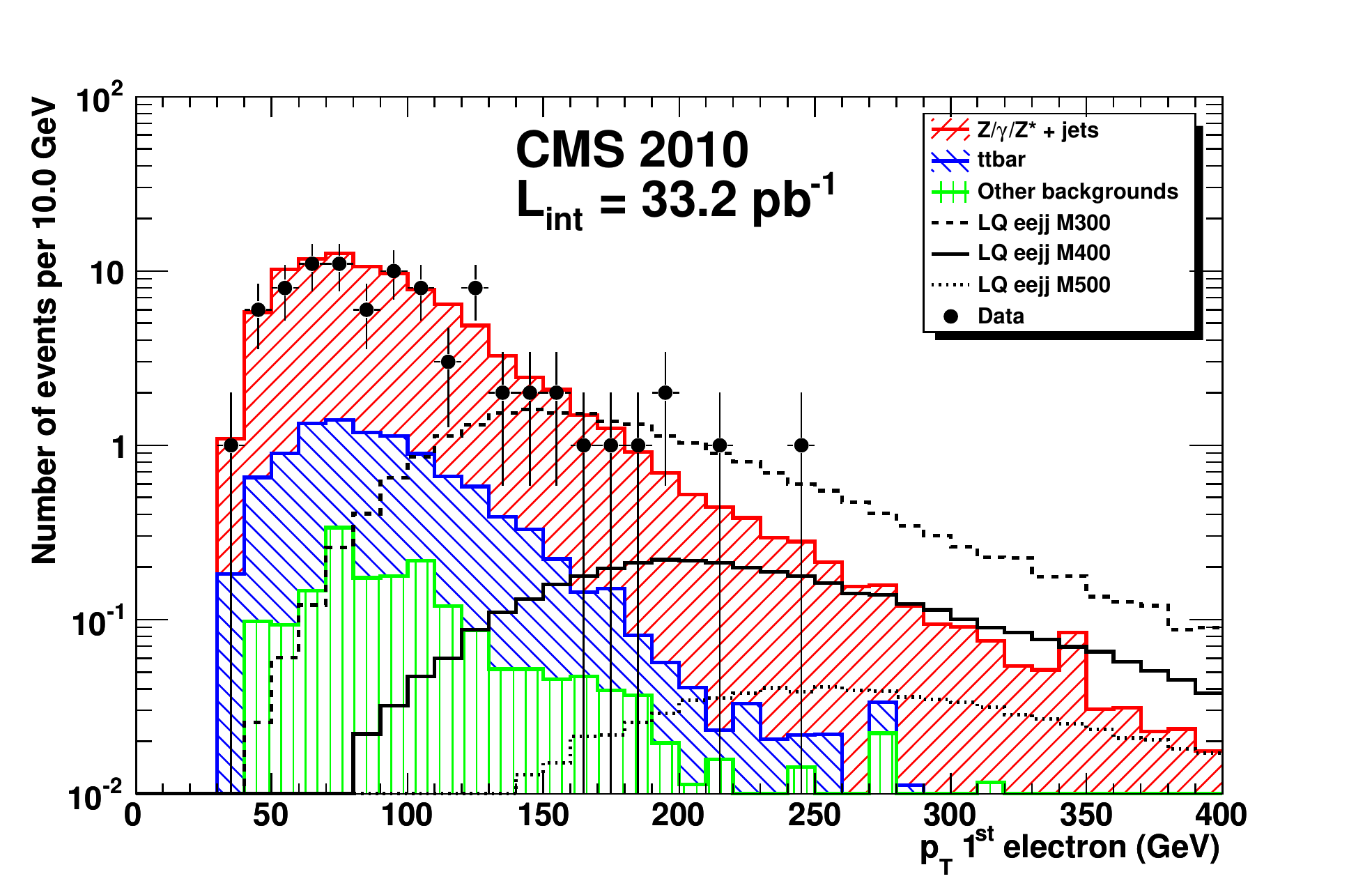}
\caption{Leading electron $p_T$ at preselection level for the $ee jj$ analysis.} \label{eeprelpte}
\end{figure}

\begin{figure}[ht]
\centering
\includegraphics[width=80mm]{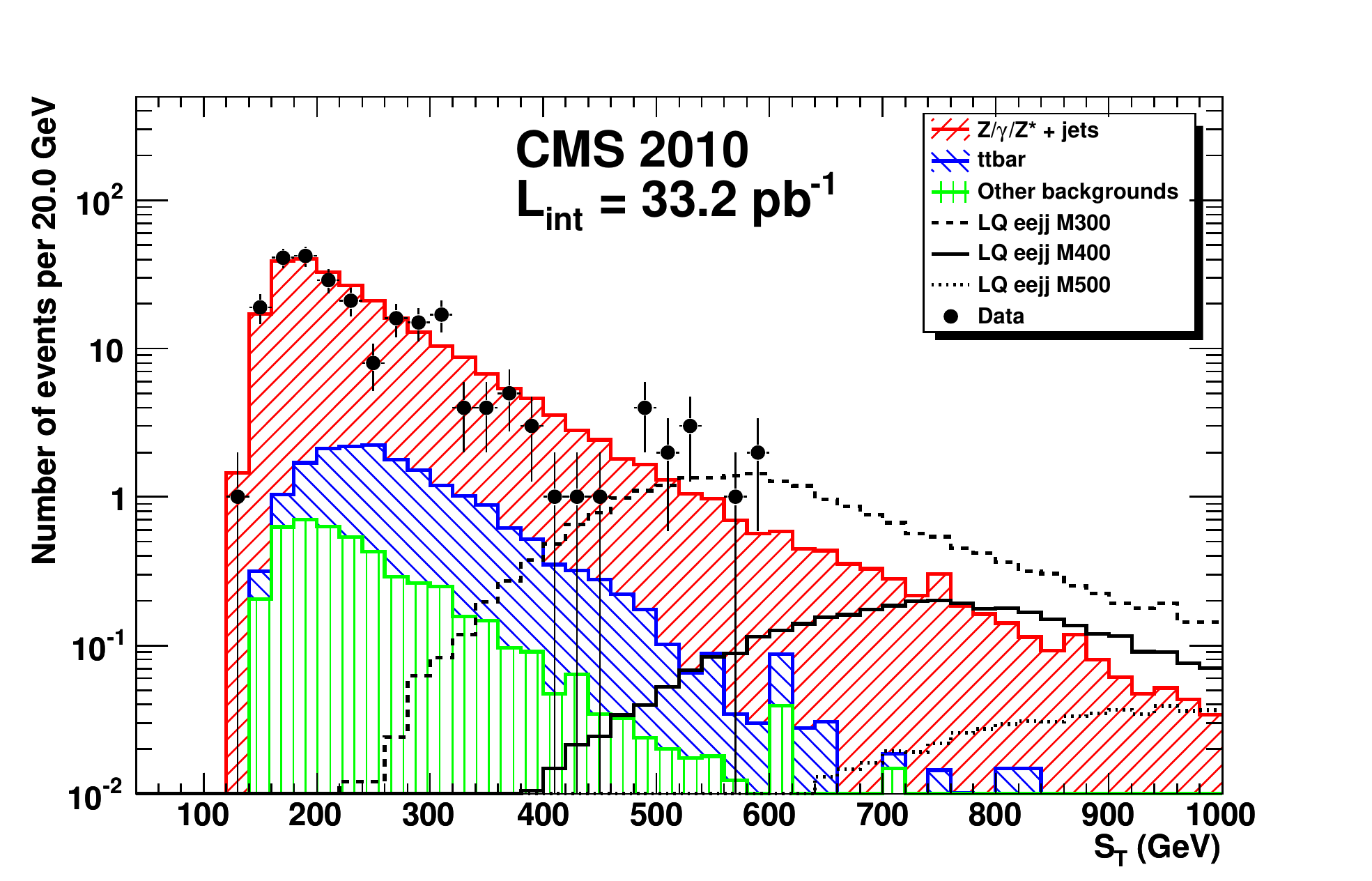}
\caption{$S_T$ at preselection level for the $ee jj$ analysis.} \label{eeprelst}
\end{figure}

The $ee jj$ analysis requires the presence of a single or double electromagnetic trigger (with an efficiency close to 100$\%$), two
or more isolated electrons with $p_T>30$ GeV and $|\eta|<2.5$, and two or more jets with $p_T>30$ GeV and $|\eta|<3.0$. In addition, at pre-selection
level, $\Delta R(e,j)>0.7$ is required together with a minimum threshold of $M_{ee}>50$ GeV and $S_T = p_T(e_1) + p_T(e_2) + p_T(jet_1) + p_T(jet_2)>250$ GeV.
Data and standard model background predictions agree well at the level of pre-selection, as shown in Fig.~\ref{eeprelpte}, and ~\ref{eeprelst}. This channel uses an integrated luminosity of 33.2 pb$^{-1}$.

\begin{figure}[ht]
\centering
\includegraphics[width=80mm]{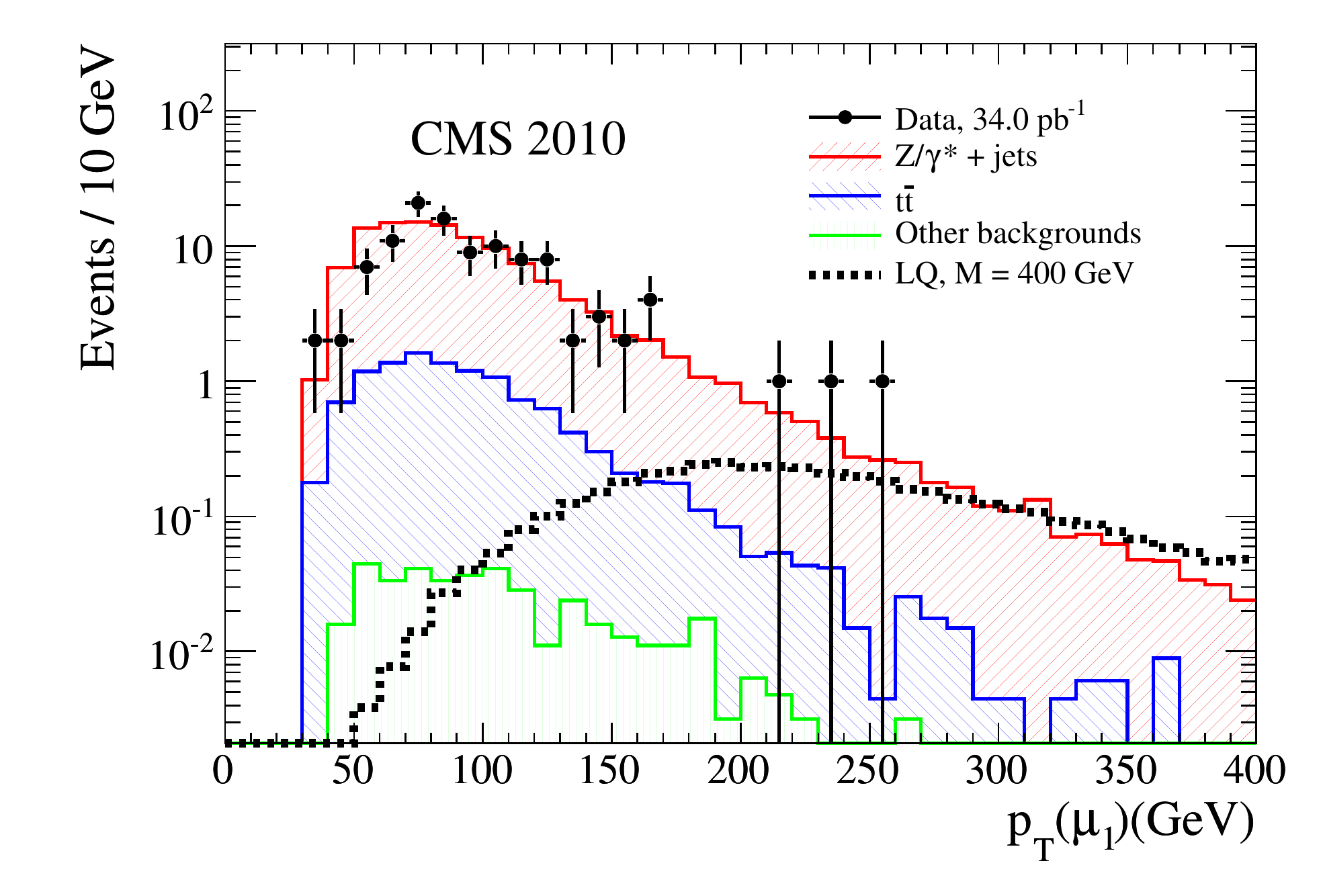}
\caption{Leading muon $p_T$ at preselection level for the $\mu\mu jj$ analysis.} \label{mmprelptm}
\end{figure}

\begin{figure}[ht]
\centering
\includegraphics[width=80mm]{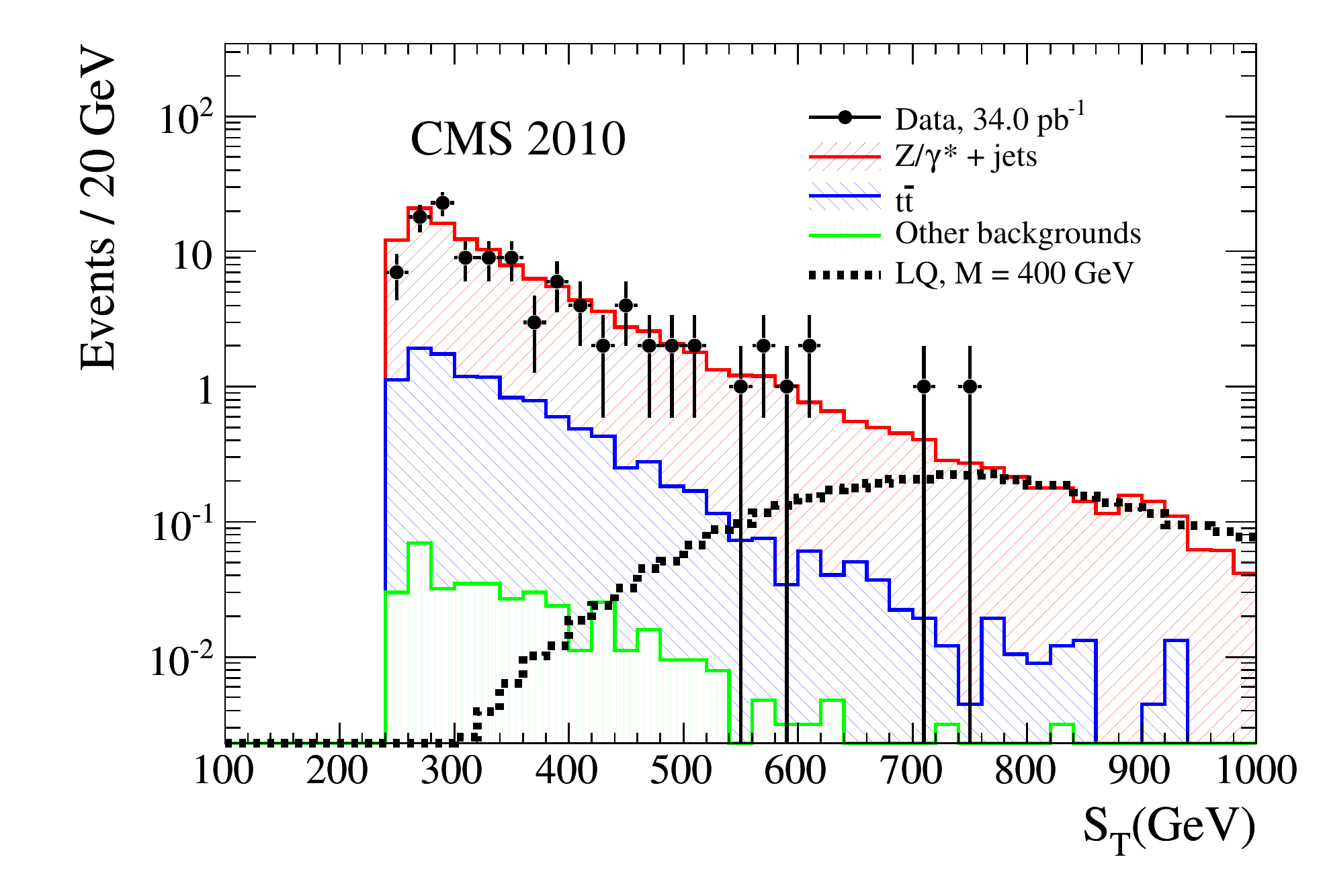}
\caption{$S_T$ at preselection level for the $\mu\mu jj$ analysis.} \label{mmprelst}
\end{figure}

Similarly, the $\mu\mu jj$ analysis starts from the requirement of a single muon trigger (with an efficiency of 99$\%$),
two or more isolated muons with $p_T>30$ GeV and $|\eta|<2.4$ (one of which must be within $|\eta|<2.1$), and two or more jets
with $p_T>30$ GeV and $|\eta|<3.0$. The two muons are required to be separated in $R$ by at least 0.3. Minimum thresholds of $M_{\mu\mu}>50$ GeV
and $S_T = p_T(\mu_1) + p_T(\mu_2) + p_T(jet_1) + p_T(jet_2)>250$ GeV complete the preselection requirements. Good agreement is observed between
data and standard model background predictions at this level (Fig.s~\ref{mmprelptm} and ~\ref{mmprelst}). This channel uses an integrated luminosity of 34.0 pb$^{-1}$.

\begin{figure}[ht]
\centering
\includegraphics[width=80mm]{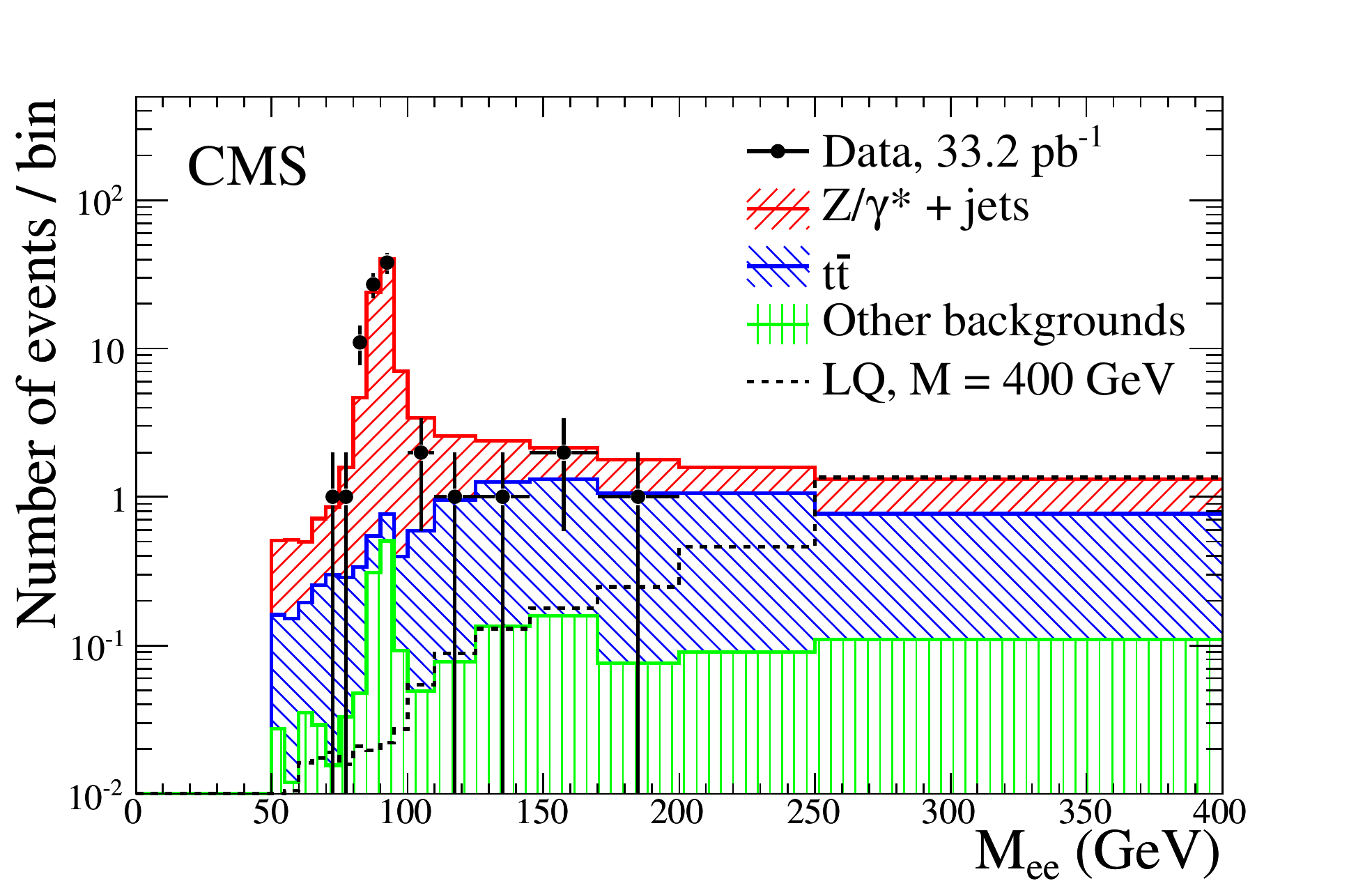}
\caption{Invariant mass of the two highest $p_T$ electrons at preselection level for the $ee jj$ analysis.} \label{eepreldimass}
\end{figure}

\begin{figure}[ht]
\centering
\includegraphics[width=80mm]{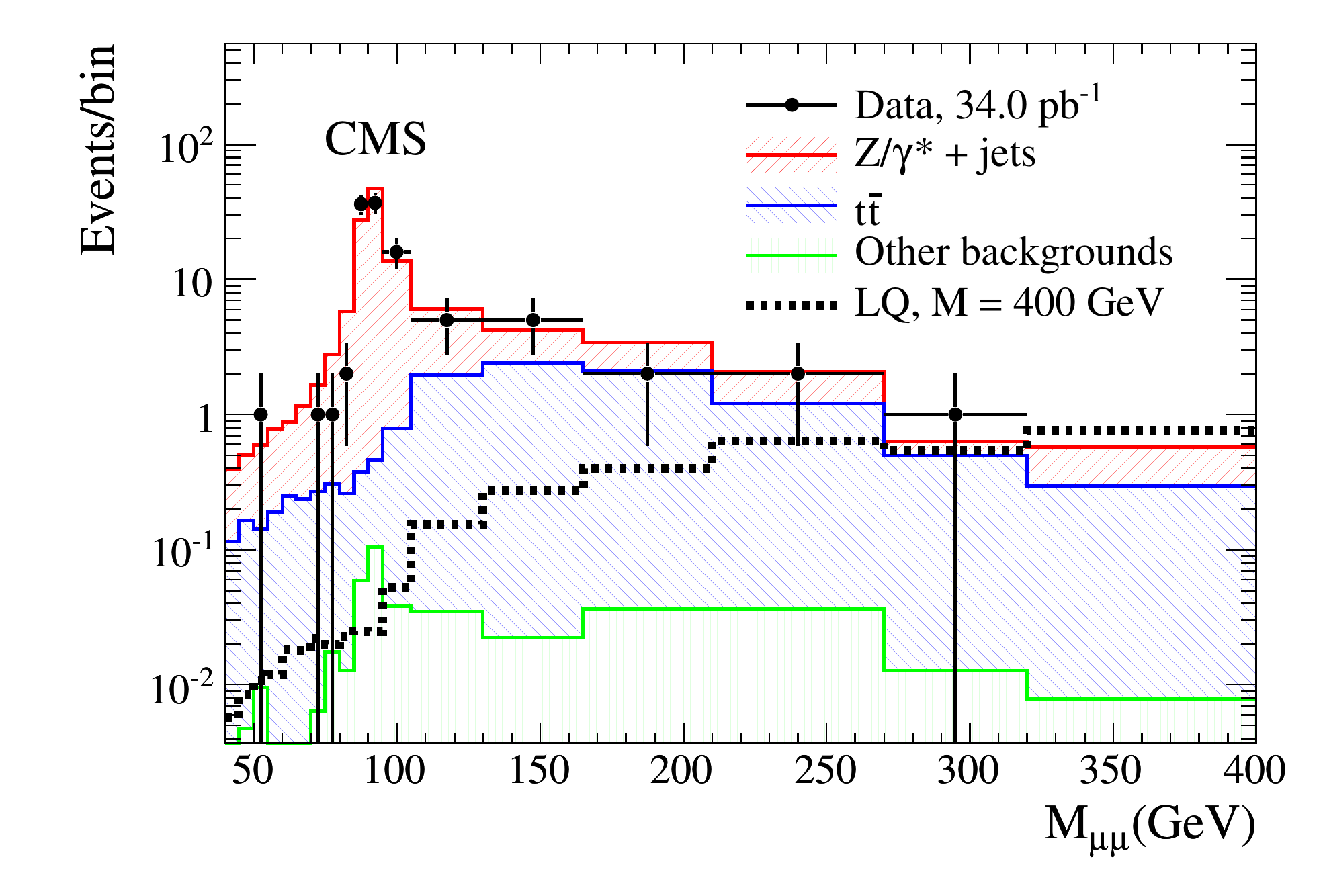}
\caption{Invariant mass of the two highest $p_T$ muons at preselection level for the $\mu\mu jj$ analysis.} \label{mmpreldimass}
\end{figure}

The major backgrounds from standard model processes to the dilepton and jets channels come from $Z/\gamma^* +$ jets and
$t\bar{t}$ production. The $Z/\gamma^*+$ contribution is determined from MC, but it is rescaled to the
data in the Z peak region at preselection. The invariant mass of the dilepton pair is shown in Fig.~\ref{eepreldimass} and in Fig.~\ref{mmpreldimass}
for the $ee jj$ and $\mu\mu jj$ channel, respectively. The normalization factors are $1.20\pm 0.14$ for $ee jj$ and
$1.28\pm 0.14$ for $\mu\mu jj$. The systematic uncertainty on the $Z/\gamma^* +$ jets background prediction in both channels comes from
the statistically dominated uncertainty in the normalization factors and from a shape uncertainty obtained by comparing
the yields of $Z/\gamma^* +$ jets MC samples generated with different renormalization and factorization
scales and matching thresholds. The $t\bar{t}$ normalization uncertainty is based on a CMS measurement of the
$t\bar{t}$ cross-section~\cite{topcms}. Smaller $W+$ jets and diboson plus jets backgrounds are determined entirely from MC and are found to be negligible.
The background contributions from QCD multijet processes are determined from data control regions and are also found to
be negligible.
At final selection, a dilepton invariant mass cut ($M_{ee}>$125 GeV and $M_{\mu\mu}>$115 GeV) is placed to suppress most of the $Z/\gamma^*+$ jets background and a $S_T$ threshold is optimized for each $M_{LQ}$ hypothesis. $S_T$ is effective at removing most of the $t\bar{t}$ background and any residual backgrounds
surviving the preselection criteria and the cut on the dilepton invariant mass. The number of data, signal MC, and background events for each of the final selection criteria optimized for different $M_{LQ}$ are listed in Table~\ref{tab:eejjsummary} and Table~\ref{tab:mumujjsummary} for the $ee jj$ and $\mu\mu jj$ channels respectively, together with the optimized $S_T$ thresholds.

\begin{table*}[htbp]
\caption{Number of $ee jj$ events for LQ signal, backgrounds,
and data after full selection with 33.2 pb$^{-1}$.
The product of signal acceptance and efficiency is also reported for different LQ
masses. The $Z/\gamma^*+$jets MC has been normalized to the data as described in the text.
Other backgrounds include $W$ + jets, di-boson, and single top.
The uncertainties are statistical.
The observed and expected 95$\%$ C.L. upper limit (u.l.) on the LQ
pair production cross section $\sigma$ are shown in the last column.}
\begin{center} \scriptsize 
\begin{tabular}{|c|cc|cccc|c|c|}
\hline\hline
$M_{\rm LQ}$    & \multicolumn{2}{|c|}{Signal Samples (MC)} & \multicolumn{4}{|c|}{Standard Model Background Samples (MC)}           & Events & Obs./Exp.\\
($S_T$~Cut)   & Selected        & Acceptance            & \multicolumn{4}{|c|}{Selected Events in}                       & in     & 95\% C.L.\\
{[GeV]}         & Events          & $\times$Efficiency    & $t\bar{t}$ + jets & $Z/\gamma^*$+ jets & Others      & All      & Data   & u.l. on $\sigma$ [pb]\\
\hline
200 ($S_T>$340) & 117.5$\pm$0.8  & 0.297$\pm$0.002 & 2.6 $\pm$0.1  & 2.0 $\pm$0.2  & 0.27$\pm$0.05 & 4.9 $\pm$0.2  & 2 & 0.441 / 0.720 \\
250 ($S_T>$400) &  43.8$\pm$0.2  & 0.380$\pm$0.002 & 1.3 $\pm$0.1  & 1.3 $\pm$0.1  & 0.14$\pm$0.02 & 2.7 $\pm$0.1  & 1 & 0.309 / 0.454 \\
280 ($S_T>$450) &  24.4$\pm$0.1  & 0.403$\pm$0.002 & 0.69$\pm$0.05 & 0.87$\pm$0.07 & 0.10$\pm$0.02 & 1.7 $\pm$0.1  & 1 & 0.305 / 0.373 \\
300 ($S_T>$470) &  17.3$\pm$0.09 & 0.430$\pm$0.002 & 0.52$\pm$0.05 & 0.75$\pm$0.07 & 0.10$\pm$0.02 & 1.4 $\pm$0.1  & 1 & 0.292 / 0.332 \\
320 ($S_T>$490) &  12.3$\pm$0.06 & 0.451$\pm$0.002 & 0.43$\pm$0.04 & 0.65$\pm$0.07 & 0.08$\pm$0.02 & 1.2 $\pm$0.1  & 1 & 0.283 / 0.305 \\
340 ($S_T>$510) &  8.88$\pm$0.04 & 0.469$\pm$0.002 & 0.32$\pm$0.04 & 0.56$\pm$0.06 & 0.08$\pm$0.02 & 0.96$\pm$0.08 & 1 & 0.278 / 0.279 \\
370 ($S_T>$540) &  5.55$\pm$0.02 & 0.496$\pm$0.002 & 0.26$\pm$0.03 & 0.47$\pm$0.06 & 0.07$\pm$0.02 & 0.80$\pm$0.07 & 1 & 0.267 / 0.254 \\
400 ($S_T>$560) &  3.55$\pm$0.02 & 0.522$\pm$0.002 & 0.20$\pm$0.03 & 0.41$\pm$0.05 & 0.06$\pm$0.02 & 0.67$\pm$0.07 & 1 & 0.257 / 0.234 \\
450 ($S_T>$620) &  1.70$\pm$0.01 & 0.539$\pm$0.002 & 0.12$\pm$0.02 & 0.28$\pm$0.05 & 0.02$\pm$0.01 & 0.42$\pm$0.06 & 0 & 0.174 / 0.210 \\
500 ($S_T>$660) & 0.868$\pm$0.003& 0.565$\pm$0.002 & 0.08$\pm$0.02 & 0.23$\pm$0.05 & 0.02$\pm$0.01 & 0.33$\pm$0.05 & 0 & 0.166 / 0.194 \\
\hline\hline
\end{tabular}
\end{center}
\label{tab:eejjsummary}
\end{table*}

\begin{table*}[htbp]
  \begin{center}
    \caption{ The $\mu\mu jj$ data event yields in 34.0 pb$^{-1}$ for different LQ mass hypotheses,
               together with the optimized $S_T$ threshold (in GeV) for each mass, background
               predictions, expected LQ signal events (S), and signal selection efficiency
               times acceptance ($\epsilon_S$). M$_{LQ}$ and $S_T$ values are listed in GeV.
               The $Z/\gamma^* \to \mu\mu + \text{jets}$ and $t\bar{t}$ contributions are
               rescaled by the normalization factors determined from data. Other backgrounds
               correspond to $VV$, $W+\text{jets}$, and
               multijet processes. Uncertainties are statistical.
             }
    \vspace{.25in}
    \scriptsize
    \begin{tabular}{| c|c c|c c c c|c|c|}\hline \hline
                        $M_{LQ}$ & \multicolumn{2}{c|}{MC Signal Samples} & \multicolumn{4}{c|}{Monte Carlo Background Samples} & Events & Obs./Exp. \\
                        ($S_T$ Cut)  & Selected & Acceptance & \multicolumn{4}{c|}{Selected Events in} & in & {$95 \%$ C.L.} \\
      $\text{[GeV]}$  & Events & $\times$ Efficiency & $t\bar{t}$ + jets & $Z/\gamma^* +\text{jets}$ & Others & All & Data& u.l. on $\sigma $ [pb]\\ \hline
200 ($S_T>310$) &160$\pm$20&0.388$\pm$0.003&4.6$\pm$0.1&4.08$\pm$0.07&0.1$\pm$0.01&8.8$\pm$0.2&5&              0.438 / 0.695   \\
225 ($S_T>350$) &89$\pm$9&0.421$\pm$0.003&3.1$\pm$0.1&2.99$\pm$0.05&0.07$\pm$0.01&6.2$\pm$0.1&3&               0.339 / 0.547  \\
250 ($S_T>400$) &51$\pm$5&0.437$\pm$0.003&1.88$\pm$0.09&1.92$\pm$0.04&0.051$\pm$0.009&3.9$\pm$0.1&3&           0.366 / 0.436  \\
280 ($S_T>440$) &28$\pm$3&0.467$\pm$0.003&1.15$\pm$0.07&1.53$\pm$0.03&0.038$\pm$0.008&2.72$\pm$0.08&3&         0.371 / 0.361  \\
300 ($S_T>440$) &21$\pm$2&0.518$\pm$0.004&1.15$\pm$0.07&1.53$\pm$0.03&0.038$\pm$0.008&2.72$\pm$0.08&3&         0.335 / 0.326  \\
320 ($S_T>490$) &14$\pm$1&0.509$\pm$0.004&0.64$\pm$0.05&1.12$\pm$0.02&0.019$\pm$0.005&1.78$\pm$0.06&2&         0.300 / 0.292  \\
340 ($S_T>530$) &9$\pm$1&0.508$\pm$0.003&0.4$\pm$0.04&0.79$\pm$0.01&0.01$\pm$0.004&1.20$\pm$0.04&1&            0.245 /0.264   \\
400 ($S_T>560$) &4.0$\pm$0.4&0.578$\pm$0.004&0.31$\pm$0.04&0.67$\pm$0.01&0.01$\pm$0.004&0.99$\pm$0.04&1&      0.219 / 0.222  \\
450 ($S_T>620$) &1.9$\pm$0.2&0.600$\pm$0.004&0.19$\pm$0.03&0.49$\pm$0.01&0.006$\pm$0.003&0.69$\pm$0.03&0&     0.153 / 0.199  \\
500 ($S_T>700$) &0.9$\pm$0.1&0.602$\pm$0.004&0.09$\pm$0.02&0.277$\pm$0.006&0.003$\pm$0.002&0.37$\pm$0.02&0&    0.152 / 0.180  \\\hline\hline
   \end{tabular}
   \label{tab:mumujjsummary}
  \end{center}
\end{table*}

\begin{figure}[ht]
\centering
\includegraphics[width=80mm]{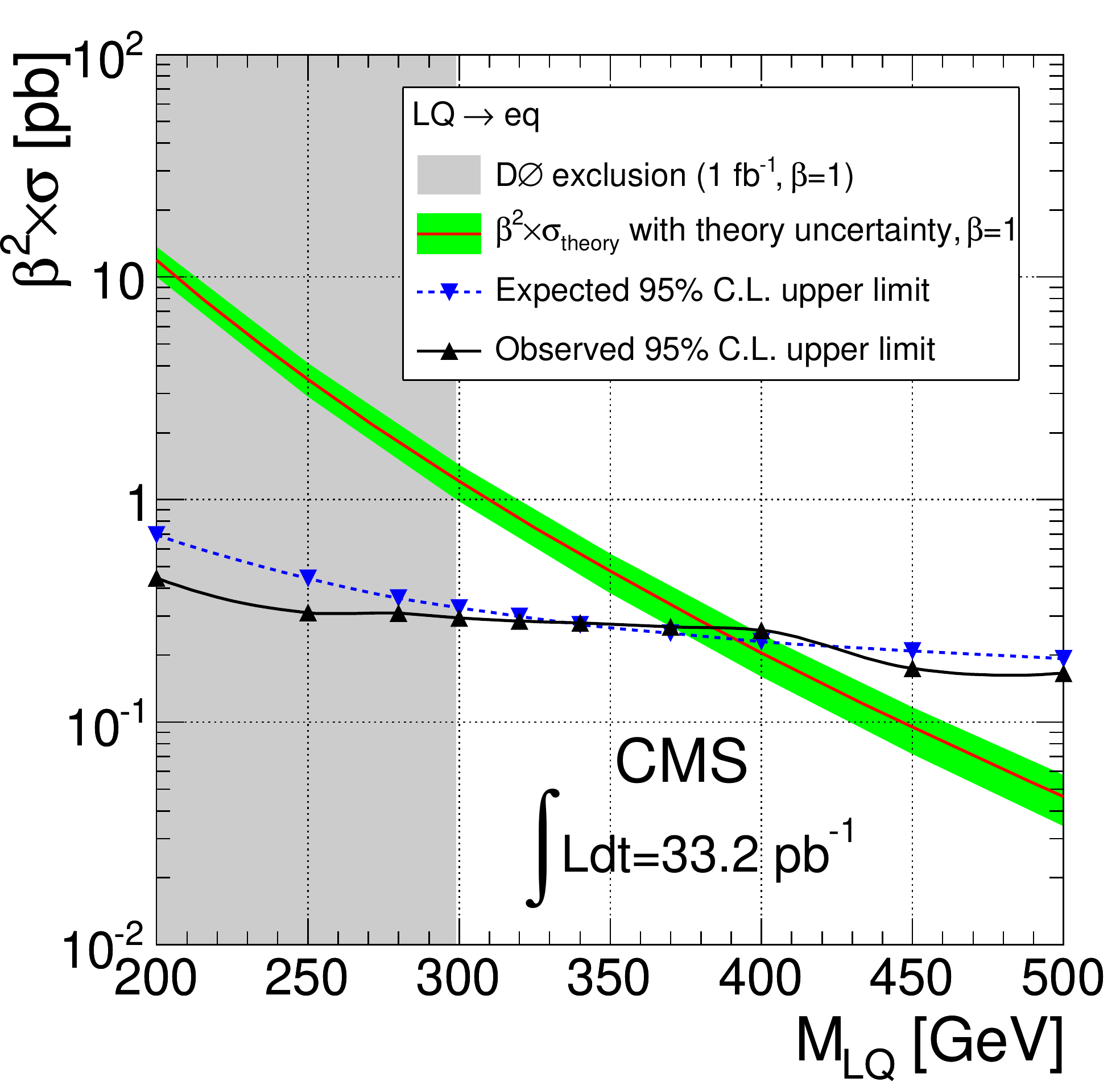}
\caption{The expected and observed 95\% C.L. upper limit on the first-generation scalar
LQ pair production $\sigma \times \beta^2$ as a function of the LQ mass
together with the NLO theoretical cross section curve. The shaded band on the theoretical values includes
CTEQ6 PDF uncertainties and the error on the LQ production cross section due
to re-normalization and refactorization scale variation.
The results correspond to 33.2~pb$^{-1}$ in the $ee jj$ channel.} \label{limit_eejj}
\end{figure}

\begin{figure}[ht]
\centering
\includegraphics[width=80mm]{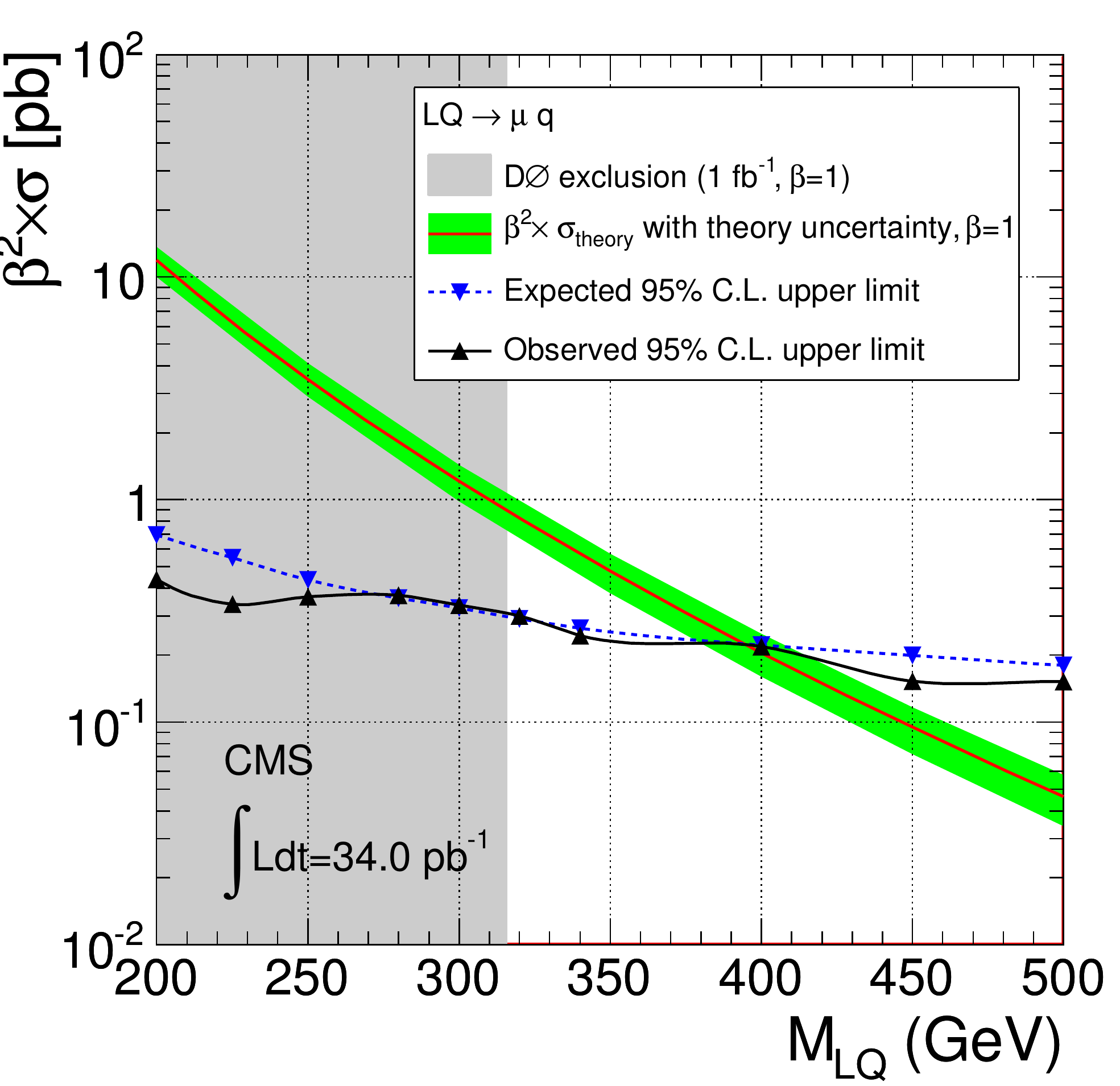}
\caption{The expected and observed 95\% C.L. upper limit on the second-generation scalar
LQ pair production $\sigma \times \beta^2$ as a function of the LQ mass
together with the NLO theoretical cross section curve. The shaded band on the theoretical values includes
CTEQ6 PDF uncertainties and the error on the LQ production cross section due
to re-normalization and refactorization scale variation.
The results correspond to 34.0~pb$^{-1}$ in the $\mu\mu jj$ channel.} \label{limit_mmjj}
\end{figure}

In absence of an excess above standard model backgrounds expectation, an upper limit on the LQ cross section is set using
a Bayesian method~\cite{bayes1, bayes} with a flat signal prior. A log-normal probability density function is used to integrate
over the systematic uncertainties. Major sources of systematic uncertainties for both channels are the uncertainties on the
determination of the luminosity, the jet energy scale, the dilepton reconstruction and identification efficiencies, and the
uncertainties associated with the normalization and modeling of the main backgrounds, $Z/\gamma^*+$jets and $t\bar{t}$.
Using Poisson statistics, a 95\% confidence level (C.L.) upper limit is obtained on
$\sigma\times\beta^2$.
This is shown in Fig.~\ref{limit_eejj} and Fig.~\ref{limit_mmjj} together with the NLO~\cite{NLO} predictions for the scalar LQ
pair production cross section. The systematic uncertainties are included in the calculation as nuisance parameters.
With the assumption that $\beta=1$, first generation and second-generation scalar leptoquarks with masses less than
384 and 394~GeV are excluded at 95\% C.L.~\cite{eejj,mmjj}. This is in agreement with the expected limits of 391 and 394~GeV.

\subsection{$e\nu jj$ channel and combination with $ee jj$}
\begin{figure}[ht]
\centering
\includegraphics[width=80mm]{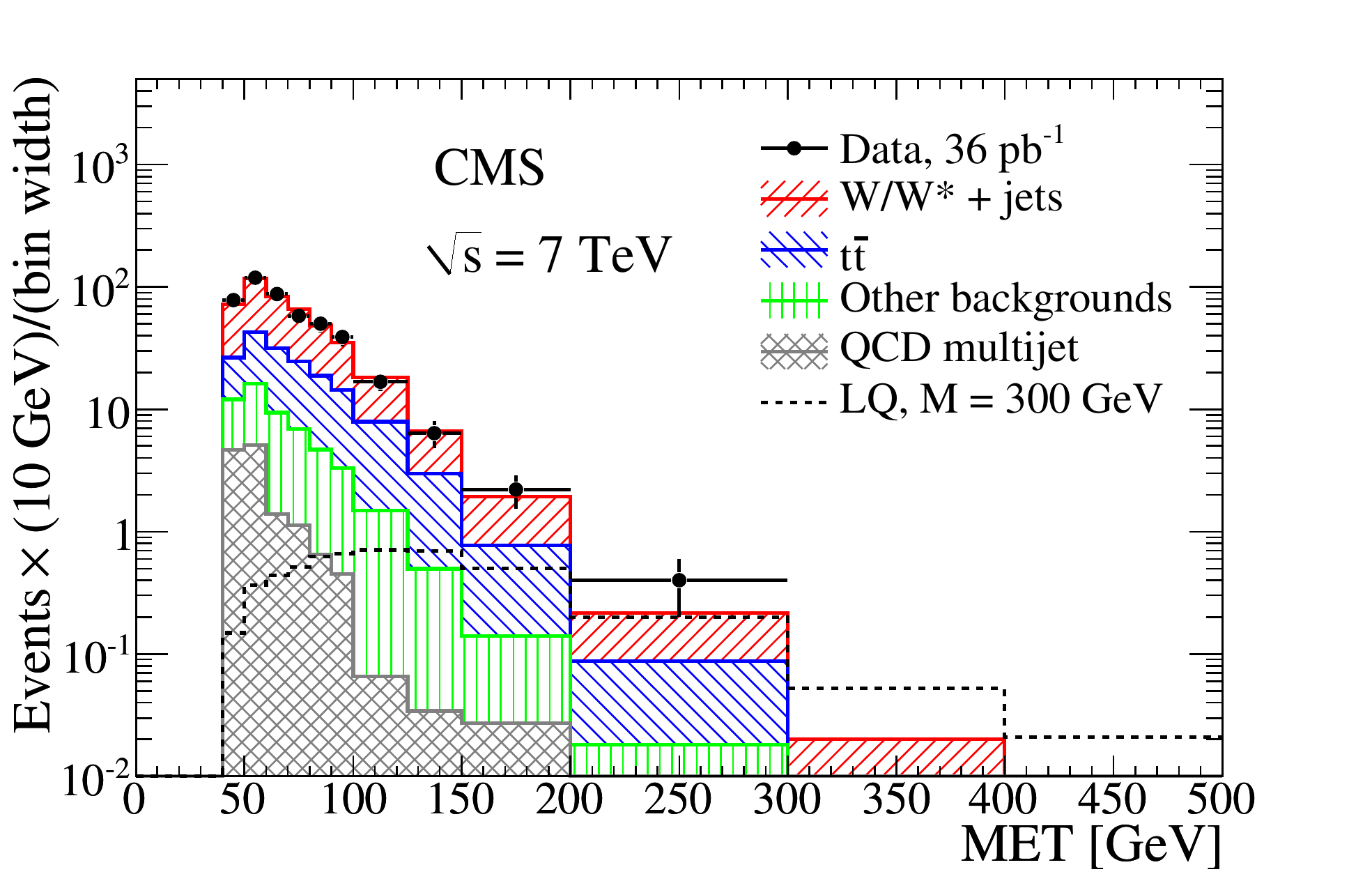}
\caption{Missing transverse energy at preselection level for the $e\nu jj$ analysis.} \label{enprelmet}
\end{figure}

The $e\nu jj$ analysis requires the presence of a single electromagnetic trigger, one
isolated electrons with $p_T>35$ GeV and $|\eta|<2.2$, and two or more jets with $p_T>30$ GeV and $|\eta|<3.0$. In addition, at pre-selection
level, $\Delta R(e,j)>0.7$ is required together with transverse missing energy, $MET>50$ GeV, a veto on the presence of muons in the event,
$S_T = p_T(e) + MET + p_T(jet_1) + p_T(jet_2)>250$ GeV, and azimuthal cuts between final state objects, $|\Delta \phi(MET,e)| >$ 0.8 and
$|\Delta \phi(MET,jet_1)| >$ 0.5 to reduce the contribution from mis-reconstructed events.
Data and standard model background predictions agree well at the level of pre-selection, as shown in
Fig.~\ref{enprelmet}. This channel uses an integrated luminosity of 36 pb$^{-1}$.


\begin{figure}[ht]
\centering
\includegraphics[width=80mm]{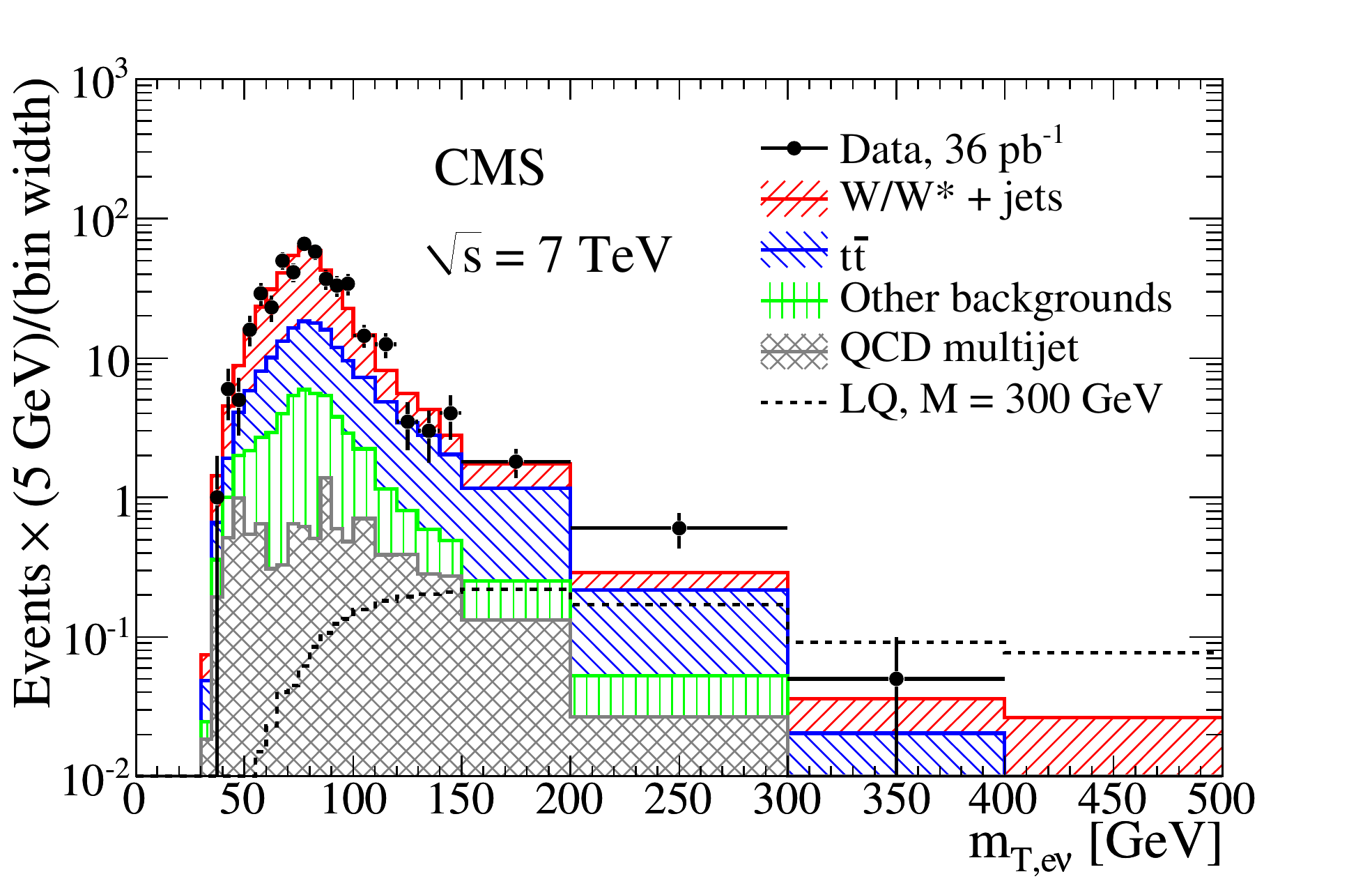}
\caption{Transverse mass of the electron and neutrino, $M_T(e\nu)$, at preselection level for the $e\nu jj$ analysis.} \label{enuprelmt}
\end{figure}

The major backgrounds from standard model processes to the $e\nu jj$ channel are $W+$ jets and
$t\bar{t}$ production. The $W+$jets contribution is determined from MC and rescaled to the
data in the 50$< M_T(e\nu) <$110 GeV region at preselection. The $M_T(e\nu)$ distribution is shown in Fig.~\ref{enuprelmt}.
The systematic uncertainty on the $W+$ jets background prediction comes from
the statistically dominated uncertainty in the normalization factors and from a shape uncertainty obtained by comparing
the yields of $W+$ jets MC samples generated with different renormalization and factorization
scales and matching thresholds. The $t\bar{t}$ normalization uncertainty is based on a CMS measurement of the
$t\bar{t}$ cross-section~\cite{topcms} and the shape uncertainty is determined from MC.
Other sources of background determined from MC are found to be negligible.
The background contributions from QCD multijet processes are determined from data control regions and are also found to be negligible.

At final selection, the transverse mass of the electron and neutrino $M_T(e\nu)$ is required to be above 125 GeV and a min$(p_T(e), MET)>$85 GeV cut is placed to suppress most of the $W+$ jets background. A $S_T$ threshold is then optimized for each $M_{LQ}$ hypothesis. $S_T$ is effective at removing most of the $t\bar{t}$ background and any residual backgrounds
surviving the preselection criteria and the $W$ veto cuts listed above. The number of data, signal MC, and background events for each of the final selection criteria optimized for different $M_{LQ}$ are listed in Table~\ref{tab:enujjsummary}, together with the optimized $S_T$ thresholds.
\begin{table*}[htbp]
\caption{Number of $e\nu jj$ events for the first generation LQ signal, backgrounds,
and data samples after the full analysis selection.
The optimum $S_T$ threshold
is reported for each LQ mass.
All uncertainties are statistical.
The product of signal acceptance and efficiency is also
reported for different LQ masses.}
\label{tab:enujjsummary}
\begin{center} \scriptsize
\begin{tabular}{|c|cc|ccccc|c|}
\hline
$M_{\text{LQ}}$  & \multicolumn{2}{c|}{MC Signal Samples} & \multicolumn{5}{c|}{MC and QCD Background Samples} & Events \\
($S_T$~cut) & Selected & Acceptance  & \multicolumn{5}{c|}{Selected Events in}  & in\\
{[GeV]}     & Events   & $\times$ Efficiency & $t\bar{t}$ + jets   & $W + $jets   & Other Bkgs   & QCD  & All Bkgs & Data\\
\hline \hline
 200 ($S_T>350$) & 34.5$\pm$0.2    & 0.161 & 3.6$\pm$0.1   & 2.2$\pm$0.3   & 0.48$\pm$0.06 & 0.20$\pm$0.04 & 6.5$\pm$0.3   & 5 \\
 250 ($S_T>410$) & 15.9$\pm$0.1  & 0.255 & 2.24$\pm$0.09 & 1.7$\pm$0.3   & 0.35$\pm$0.05 & 0.18$\pm$0.05 & 4.4$\pm$0.3   & 3 \\
 280 ($S_T>460$) & 9.54$\pm$0.05   & 0.291 & 1.43$\pm$0.08 & 1.2$\pm$0.2   & 0.29$\pm$0.05 & 0.14$\pm$0.04 & 3.1$\pm$0.2   & 3 \\
 300 ($S_T>490$) & 6.89$\pm$0.03   & 0.317 & 1.09$\pm$0.07 & 1.0$\pm$0.2   & 0.27$\pm$0.05 & 0.14$\pm$0.04 & 2.5$\pm$0.2   & 2 \\
 320 ($S_T>520$) & 5.03$\pm$0.02   & 0.339 & 0.75$\pm$0.05 & 0.8$\pm$0.2   & 0.22$\pm$0.05 & 0.13$\pm$0.04 & 1.9$\pm$0.2   & 2 \\
 340 ($S_T>540$) & 3.73$\pm$0.02   & 0.364 & 0.65$\pm$0.05 & 0.7$\pm$0.2   & 0.20$\pm$0.05 & 0.12$\pm$0.04 & 1.6$\pm$0.2   & 2 \\
 370 ($S_T>570$) & 2.40$\pm$0.01 & 0.396 & 0.50$\pm$0.04 & 0.6$\pm$0.1   & 0.18$\pm$0.04 & 0.08$\pm$0.03 & 1.3$\pm$0.2   & 1 \\
 400 ($S_T>600$) & 1.57$\pm$0.01 & 0.426 & 0.34$\pm$0.04 & 0.5$\pm$0.1   & 0.17$\pm$0.04 & 0.08$\pm$0.03 & 1.1$\pm$0.1   & 1 \\
 450 ($S_T>640$) & 0.797$\pm$0.003 & 0.467 & 0.26$\pm$0.03 & 0.4$\pm$0.1   & 0.13$\pm$0.04 & 0.08$\pm$0.04 & 0.9$\pm$0.1   & 0 \\
 500 ($S_T>670$) & 0.417$\pm$0.001 & 0.500 & 0.18$\pm$0.03 & 0.4$\pm$0.1   & 0.12$\pm$0.04 & 0.08$\pm$0.04 & 0.8$\pm$0.1   & 0 \\
\hline
\end{tabular}
\end{center}
\end{table*}

\begin{figure}[ht]
\centering
\includegraphics[width=80mm]{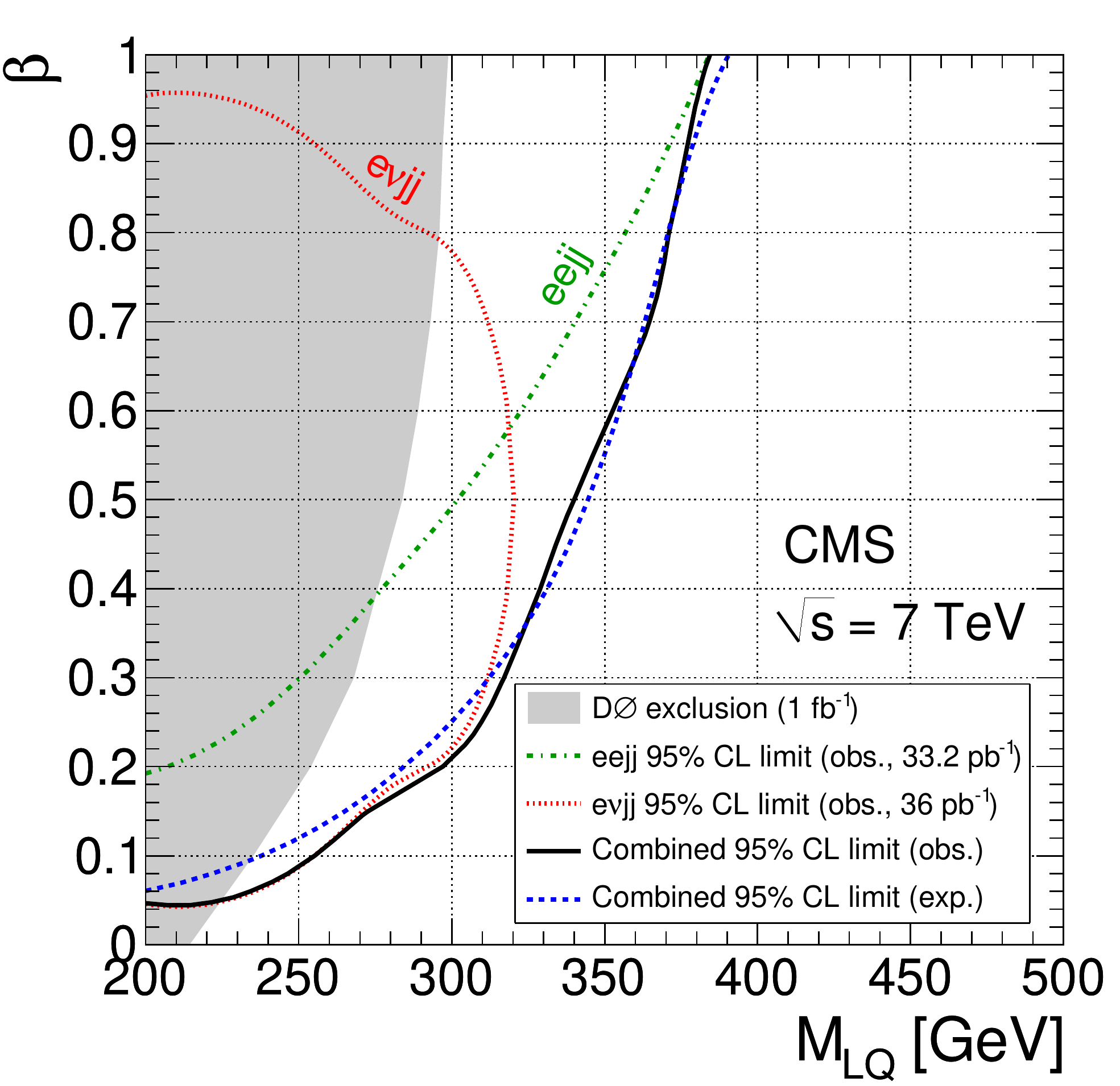}
\caption{The combined expected and observed 95\% C.L. mass limit for first-generation scalar
leptoquarks as a function of $\beta$. Individual channel limits ($ee jj$ and $e\nu jj$) are also shown.} \label{eecombbeta}
\end{figure}

In absence of an excess above standard model backgrounds expectation, an upper limit on the LQ cross section is set using
a Bayesian method similar to the one used in the $ee jj$ analysis. The $e\nu jj$ channel has maximum
sensitivity for $\beta = 0.5$, and also offers sensitivity for lower values of $\beta$, where the $ee jj$ channel
quickly runs out of sensitivity due to the quadratic behavior of the branching fraction of the LQ pair into the above channel.
Thus, combining it with the $ee jj$ channel allows one to extend the sensitivity of the search in the intermediate
$\beta$ range compared to either of the two analysis considered separately.
The two channels are combined using the same Bayesian approach used to set the individual limits. Assuming a flat prior for
the  signal cross section, the expected signal yield in the $ee jj$ [$e\nu jj$] channel is found by
multiplying the cross section by the signal efficiency, integrated luminosity, and the branching fraction
$\beta^2$ [$2\beta(1-\beta)$]. Since the major sources of systematic uncertainty between the two channels are
correlated, a combined likelihood is constructed from the individual likelihoods for the two channels.
The 95\% C.L. mass limit as a function of $\beta$ is shown in Fig.~\ref{eecombbeta}, along with the individual limits
from the two channels. With the assumption that $\beta=0.5$, first generation scalar leptoquarks with masses less than
340~GeV are excluded at 95\% C.L.~\cite{enjj}.

\subsection{Conclusions and outlook}
Searches for pair production of first and second generation scalar leptoquarks have been performed at CMS
in the final states with: two charged leptons and two jets, one charged lepton (electron), missing transverse energy
and two jets, with the full 2010 statistics. In the absence of a signal we exclude first generation LQ with masses below 384 GeV
($\beta =1$), second generation LQ with masses below 394 GeV ($\beta =1$), and first generation LQ with masses
below 340 GeV ($\beta = 0.5$). The $\beta = 1$ results were the most stringent limits at the time of publication.
The $\beta = 0.5$ and $\beta = 1$ results have been combined and have been submitted for publication.
The analysis of the 2011 data is ongoing.


\end{document}